\documentclass[nonacm,sigconf]{aamas}

\usepackage{balance} %

\usepackage{svg}
\usepackage{epsdice}

\usepackage{amsmath,amsfonts,bm,bbm,xfrac}

\def\eqref#1{equation~(\ref{#1})}
\def\Eqref#1{Equation~(\ref{#1})}

\def\plaineqrefp#1{(\ref{#1})}

\def\Algref#1{Algorithm~\ref{#1}}

\def\Twoalgref#1#2{Algorithms \ref{#1} and \ref{#2}}

\def\appref#1{Appendix~\ref{#1}}

\def\1{\bm{1}}

\usepackage{cancel}

\usepackage{listings}
\usepackage{enumerate}

\def\va{{\bm{a}}}

\def\vx{{\bm{x}}}

\DeclareMathAlphabet{\mathsfit}{\encodingdefault}{\sfdefault}{m}{sl}
\SetMathAlphabet{\mathsfit}{bold}{\encodingdefault}{\sfdefault}{bx}{n}

\DeclareMathOperator*{\argmax}{arg\,max}

\usepackage{amsthm}

\makeatletter
\newtheorem*{rep@theorem}{\rep@title}
\newcommand{\newreptheorem}[2]{%
\newenvironment{rep#1}[1]{%
 \def\rep@title{#2 \ref{##1}}%
 \begin{rep@theorem}}%
 {\end{rep@theorem}}}
\makeatother

\newreptheorem{theorem}{Theorem}

\newreptheorem{proposition}{Proposition}

\newreptheorem{lemma}{Lemma}
\newtheorem{assumption}{Assumption}

\def\br{{BR}}

\usepackage{algorithm,algorithmic}

\usepackage{multirow}

\definecolor{green}{rgb}{0.194, 0.605, 0.316}
\definecolor{blue}{rgb}{0.191, 0.505, 0.742}
\definecolor{orange}{rgb}{0.8, 0.33, 0.} %
\definecolor{cobalt}{rgb}{0.0, 0.28, 0.67} %
\definecolor{egred}{rgb}{1.0, 0.25, 0.25}
\definecolor{codegreen}{rgb}{0,0.6,0}
\definecolor{codegray}{rgb}{0.5,0.5,0.5}
\definecolor{codepurple}{rgb}{0.58,0,0.82}
\definecolor{backcolour}{rgb}{0.95,0.95,0.92}

\setcopyright{ifaamas}
\acmConference[AAMAS '25]{Proc.\@ of the 24th International Conference
on Autonomous Agents and Multiagent Systems (AAMAS 2025)}{May 19 -- 23, 2025}
{Detroit, Michigan, USA}{A.~El~Fallah~Seghrouchni, Y.~Vorobeychik, S.~Das, A.~Nowe (eds.)}
\copyrightyear{2025}
\acmYear{2025}
\acmDOI{}
\acmPrice{}
\acmISBN{}

\acmSubmissionID{<<EasyChair submission id>>}

\title[Nash Equilibria via Stochastic Eigendecomposition]{Nash Equilibria via Stochastic Eigendecomposition}

\author{Ian Gemp}
\affiliation{
  \institution{Google DeepMind}
  \city{London}
  \country{United Kingdom}}
\email{imgemp@google.com}

\begin{abstract}
This work proposes a novel set of techniques for approximating a Nash equilibrium in a finite, normal-form game. It achieves this by constructing a new reformulation as solving a parameterized system of multivariate polynomials with tunable complexity. In doing so, it forges an itinerant loop from game theory to machine learning and back. We show a Nash equilibrium can be approximated with purely calls to stochastic, iterative variants of singular value decomposition and power iteration, with implications for biological plausibility. We provide pseudocode and experiments demonstrating solving for all equilibria of a general-sum game using only these readily available linear algebra tools.
\end{abstract}


\newcommand{\BibTeX}{\rm B\kern-.05em{\sc i\kern-.025em b}\kern-.08em\TeX}

\begin{document}

\pagestyle{fancy}
\fancyhead{}

\maketitle

\section{Introduction}

Nash equilibrium (NE) is the central solution concept for finite, normal-form games. Unfortunately, unless PPAD $\subseteq$ P, no fully polynomial time algorithm exists to approximate it in generic, general-sum games~\citep{daskalakis2009complexity,chen2006settling,daskalakis2013complexity}. Nevertheless, it is important to develop a variety of techniques, for instance, tailored to different restricted game classes or honed to select out equilibria with particular properties.

No-regret or gradient-based approaches are particularly lightweight and have been effective in certain applications although they come with no NE-convergence guarantees beyond $2$-player, zero-sum~\citep{facchinei2007finite,gordon2008no,blackwell1956analog,vlatakis2020no}. Homotopy methods have been designed to select out specific equilibria~\citep{lemke1964equilibrium,govindan2003global,govindan2004computing,harsanyi1988general,perolat2020poincar}, some with the additional characteristic of modelling players with bounded rationality~\citep{mckelvey1995quantal,mckelvey1998quantal,turocy2005dynamic,eibelshauser2019markov,gemp2022sample}. The primary measure of approximation for NEs is \emph{exploitability}, the maximum any player can gain by deviating from an approximate equilibrium profile. Hence, there exists a line of work that directly attempts to minimize exploitability using an optimization formulation~\citep{shoham2008multiagent,sandholm2005mixed,gemp2023approximating}. Similarly, the property that no player can gain by deviating can be viewed as a constraint. Constraint satisfaction approaches were found to be effective empirically~\citep{porter2008simple}; other methods also take a search-tree approach~\citep{berg2017exclusion,gemp2023approximating}.
\begin{figure}[ht!]
    \centering
    \includegraphics[width=0.45\textwidth]{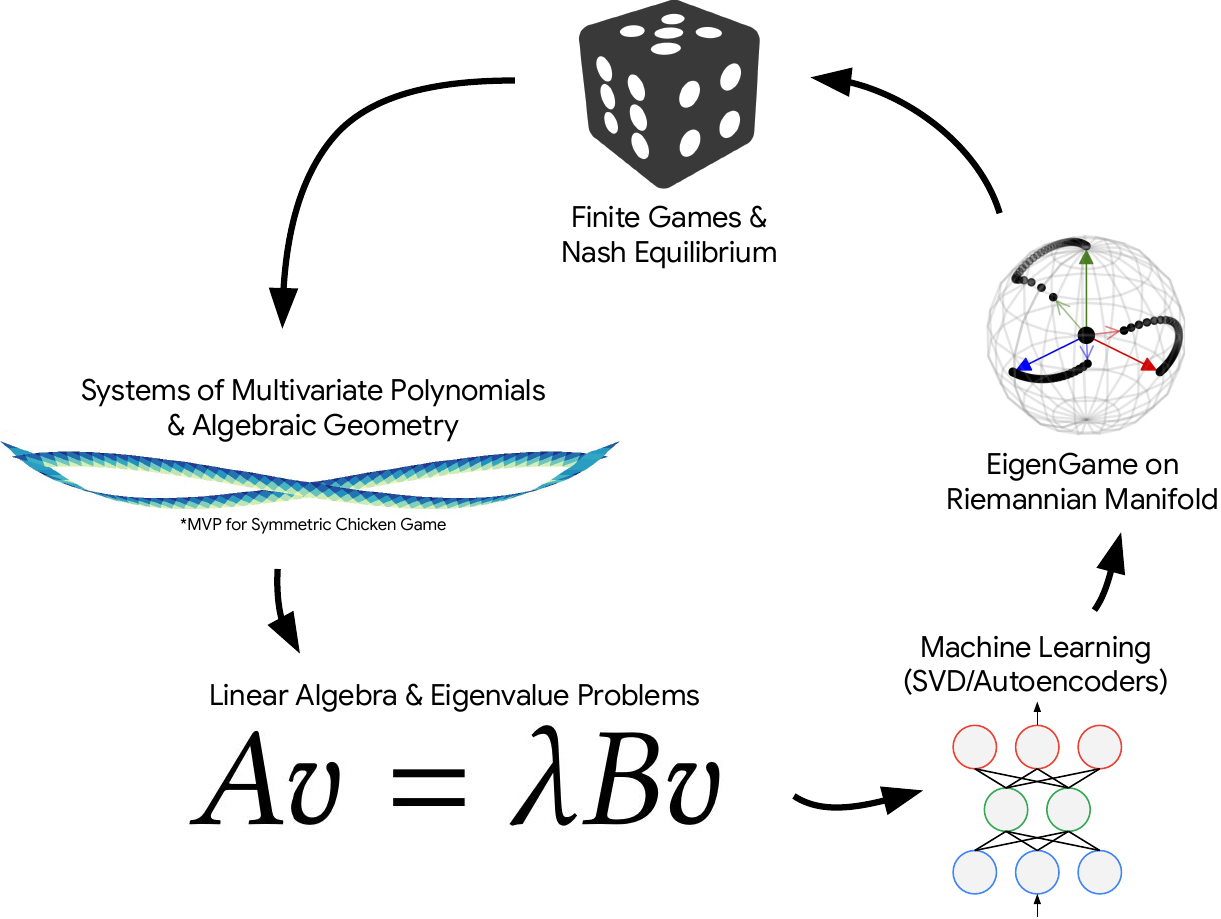}
    \caption{(An Itinerant Loop) We identify a series of bridges\textemdash some new, some old, and some a mix\textemdash that connects the problem of approximating Nash equilibria in normal-form games to problems in algebraic geometry, linear algebra, machine learning, and back again to finite games (albeit not precisely where the loop began).}
    \label{fig:itinerant_loop}
\end{figure}
The boundaries between these families are not strict: search-based approaches mix with optimization, gradient based with homotopy-based approaches, etc.

Of particular relevance to this work is the family of approaches that view the NE problem (NEP) as solving for the common roots of a system of multivariate polynomials subject to some constraints.
For instance, an NEP can be represented as a polynomial complementarity problem (PCP)~\citep{wilson1971computing}: for each player, either an action is played with zero probability at an NE or it achieves the maximum payoff possible at the NE. Probabilities can be recovered from the unconstrained solutions to the PCP via normalization.
An NEP can also be represented as a multivariate polynomial problem (MVP) \emph{with simplex constraints}~\citep[Section 6.3]{sturmfels2002solving}.
Assuming a fully-mixed NE (i.e., one that lies in the interior of the simplex) exists, it can be solved for via a similar approach as above.
\citet{sturmfels2002solving} demonstrates solving for these equilibria via PHCpack~\citep{verschelde1999algorithm}, a software package to solve polynomial systems by homotopy continuation methods. Unsurprisingly, these and other methods~\citep{li1997numerical} have parallels with the homotopy methods for approximating Nash equilibria mentioned above.

Recently, renewed interest in numerical methods for solving polynomial systems via eigendecompositions has surged. While knowledge of these techniques has existed for some time, the goal of new research is to make these techniques more accessible to non-experts~\citep{dreesen2012back,williams2010solving} as well as further developing the techniques~\citep{vermeersch2023block}. These techniques generally assume access to linear algebra routines that make few assumptions on the matrices of interest. Contrast that with machine learning (ML)~\citep{allen2017first}, where singular value decomposition (SVD) has received the bulk of the community's interest, and whose underlying eigenvalue problem assumes a symmetric, positive semi-definite matrix. The focus here has been scaling to large matrices and reducing memory requirements by leveraging stochastic access to matrix entries.

In this work, we develop a novel formulation of the approximate Nash equilibrium problem as a multivariate polynomial problem. Central to this formulation is the regularization of the game with Tsallis entropy~\citep{gemp2022sample}. Similar to other frameworks of bounded rationality such as logit equilibria~\citep{mckelvey1995quantal}, we can show this Tsallis regularized game is equivalent to a game where the log of the payoffs are perturbed by Gumbel($0, \tau$) noise (see \appref{app:gumbel}). Critically, given an approximate equilibrium $x^{(\tau)}$ of the transformed game, we also provide bounds on the exploitability of $x^{(\tau)}$ as measured in the original game similar to prior work~\citep{gemp2023approximating}.

Our MVP formulation is different from prior work and exhibits a distinct advantage. Whereas prior work results in MVPs that are at least quadratic, an NP-hard problem~\citep{courtois2002solving}, ours results in an MVP that is linear for the class of $2$-player, general-sum games and for a specific setting of the Tsallis entropy parameter $\tau$.
Therefore, this new formulation can provide fast approximate solutions to $2$-player, general-sum games via a simple least squares solver.

For the wider class of $N$-player, general-sum games, we show how to solve for NEs using only access to stochastic singular value decomposition (SVD) and power iteration~\citep{golub2013matrix}.

Lastly, given SVD's recent reformulation as a Nash equilibrium problem, specifically an EigenGame~\citep{gempeigengame}, we explain a mapping from the original NE problem to a sequence of NE problems on larger, albeit strategically simpler games (see Figure~\ref{fig:itinerant_loop}). All but the final step of our approach that uses power iteration can be interpreted as a new normal-form game thereby establishing a tree of NE problems (see Figure~\ref{fig:tree_of_games}).

\section{Background}\label{sec:bgrnd}

We first review the mathematics of finite games.
A normal-form game is a tuple $\langle N, \mathcal{A} = (\bigtimes_{i=1}^N \mathcal{A}_i), u=(u_1, \ldots, u_N)\rangle$ where $N \in \mathbb{N}$ is the number of players, $\mathcal{A}_i$ is the finite set of actions available to player $i \in [N]$, and $u_i: \mathcal{A} \rightarrow \mathbb{R}$ is player $i$'s utility function. Player's may randomize over their strategy sets, called \emph{mixed} strategies: $\mathcal{X}_i = \Delta^{\vert \mathcal{A}_i \vert - 1}$. Their utility functions are naturally extended to this domain using expected value: $u_i: \mathcal{X} = (\bigtimes_{i=1}^N \mathcal{X}_i) \rightarrow \mathbb{R} = \mathbb{E}_{\va \sim \vx}[u_i(\va)]$ where $\vx \in \mathcal{X}$ and $\va \in \mathcal{A}$. For the sake of clarity, it is worth expanding out this expectation:
\begin{align}
    u_i(\vx) &= \sum_{\va \in \mathcal{A}} u_i(\va) \prod_{j=1}^N x_{j,a_j}
\end{align}
where $x_{j,a_j}$ denotes player $j$'s probability of playing their component of the action profile $\va$ under their strategy $x_j$. From this formula, it becomes clear that player $i$'s utility is a multivariate polynomial in $\vx$. In fact, it is written as a weighted sum of monomials in $\vx$, each of degree $N$. In addition, player $i$'s gradient with respect to its strategy $x_i$, denoted $\nabla^i_{x_i}$, is clearly polynomial as well. This observation provides the foundation for prior connections between the Nash equilibrium problem and MVPs.

A player's gradient can be represented more concisely in Einstein notation. Let $U^{(i)} \in \mathbb{R}^{\bigtimes_{j=1}^N \vert \mathcal{A}_j \vert}$ be player $i$'s utility function in tensor-form such that $u_i(\va)$ can be retrieved by simply indexing this tensor as $U^{(i)}(a_1, \ldots, a_N)$. Then player $i$'s expected utility and gradient can be written respectively as:
\begin{align}
    u^i_{x_i} &= U^{(i)}_{1 \ldots N} x_1 \ldots x_N \label{eqn:ui_einstein}
    \\ \nabla^i_{x_i} &= U^{(i)}_{1 \ldots N} \ldots x_{i-1} x_{i+1} \ldots \label{eqn:gradi_einstein}
\end{align}
where $\ldots x_{i-1} x_{i+1} \ldots$ indicates all strategies except player $i$'s.

\subsection{NE with Tsallis Entropy}\label{sec:bgrnd:ne_tsallis}

There exists a rich history of solving equilibria by instead solving a curricula of transformed games, finishing with solving the original game in the limit. This approach appears under the family of Tikhonov regularization techniques in variational inequality theory~\citep{qi1999tikhonov} where the focus is on continuous strategy sets, but has applications to two-player, zero-sum normal form games. A separate line of research focuses on homotopy methods in finite (discrete strategy set) games~\citep{harsanyi1988general}. Of those, the curriculum (or continuum) of quantal response equilibria is most famous~\citep{mckelvey1995quantal,mckelvey1998quantal}. In this case, players are given a bonus to their payoff for selecting high-entropy mixed strategies: $u^{\tau}_i(\vx) = u_i(\vx) + \tau S(x_i)$ where $S$ stands for Shannon entropy. Note that $S$ is strictly concave in $x_i$, and so Nash equilibria necessarily exist in these games according to theory on $n$-player concave-utility games~\citep{rosen1965existence}; these Nash equilibria of the transformed game are precisely the quantal response equilibria defined at temperature $\tau$, denoted QRE($\tau$).

Inspired by this, prior work~\citep{gemp2022sample} considered swapping Shannon for Tsallis entropy $H^{\tau}(x_i) = \frac{\gamma}{\tau+1}(1 - \sum_l x_{il}^{\tau+1})$ weighted by $\gamma$ with $\tau \in [0,1]$. $H^{\tau}$ is also concave, and so the same NE existence property holds.

\begin{assumption}[Normalized Payoffs]\label{assump:u_01}
$u_i(\va) \in (0, 1]$ for every player $i$ and joint action $\va$.
\end{assumption}

In particular,~\citet{gemp2022sample} recognized that when Assumption~\ref{assump:u_01} holds and $\gamma = ||\nabla^i_{x_i}||_{1/\tau}$ in this transformed game\footnote{The $q$-norm of a vector $v$ is $\big(\sum_i (v_i)^{q}\big)^{\sfrac{1}{q}}$.}, each player's gradient $\nabla^{i\tau}_{x_i}$ and best response $\br_i^{\tau}$ are as follows:
\begin{align}
    \nabla^{i\tau}_{x_i} &= \nabla_{x_i} u_i^{\tau}(\boldsymbol{x}) = \nabla^i_{x_i} - \gamma x_i^{\tau} \label{eqn:grad}
    \\ \br^{\tau}_i &= \frac{(\nabla^i_{x_i})^{1/\tau}}{\sum_l (\nabla^i_{il})^{1/\tau}} \in \Delta^{\vert \mathcal{A}_i \vert -1}. \label{eqn:br}
\end{align}
\Eqref{eqn:grad} will form the basis for constructing our multivariate polynomial formulation of the Nash equilibrium problem later.

\subsection{Projected Gradients}

\begin{figure}[ht!]
    \centering
    \includegraphics[width=0.3\textwidth]{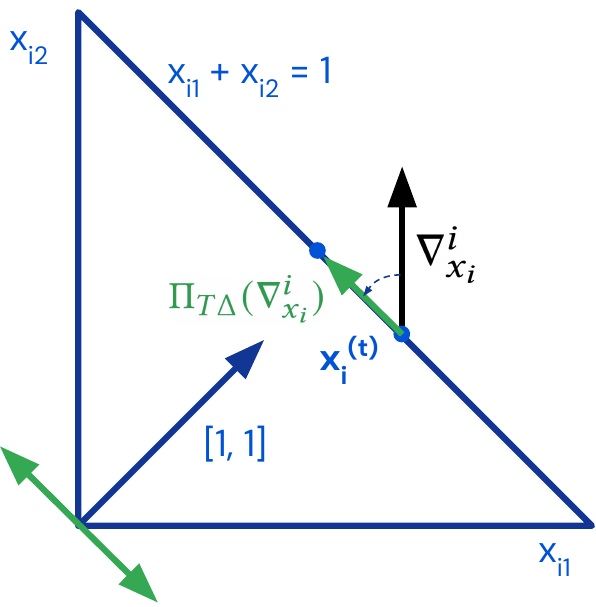}
    \caption{(Projected Gradients) A player's gradient in a normal-form game $\nabla^i_{x_i} \in \mathbb{R}^{\vert \mathcal{A}_i \vert = 2}$ can point \emph{off} the simplex and need not have zero length at an interior Nash equilibrium. In contrast, the gradient projected onto the tangent space of the simplex, \textcolor{green}{$\Pi_{T\Delta}(\nabla^i_{x_i})$}, \emph{does} necessarily have zero length for all interior Nash equilibria and is the natural notion of ``movement'' on the simplex.}
    \label{fig:proj_grad}
\end{figure}

Our MVP formulation conceptually reduces to finding strategy profiles where every player's gradient is zero. However, two modifications are necessary beyond this naive viewpoint. First, a player's gradient is not necessarily zero even at an interior NE. Consider the standard $2$-player, rock-paper-scissors game except where payoffs are offset by $+1$ (constant offsets do not affect the location of NEs). In this case, each player's gradient will be the ones vector at the NE (uniform profile), not the zeros vector. To rectify this issue, we can consider a player's ``projected gradient'' which computes the component of the player's gradient that lies in the tangent space of the simplex; the projected gradient is necessarily zero at an interior NE~\citep{gemp2023approximating} (see Figure~\ref{fig:proj_grad}). The formula for projecting the gradient is a linear operation that simply subtracts the mean value from each player's gradient vector:
\begin{align}
    \Pi_{T\Delta}\big( \nabla^{i\tau}_{x_i} \big) &= [I - \frac{1}{\vert \mathcal{A}_i \vert} \mathbf{1} \mathbf{1}^\top] \nabla^{i\tau}_{x_i}. \label{eqn:projgrad}
\end{align}

Second, in a generic game, a player's gradient might not be zero at an NE; for example, a player's gradient can be nonzero at a pure NE as in the standard prisoner's dilemma. Tsallis entropy regularization is used to ensure all NEs occur in the interior of $\mathcal{X}$, and hence all exhibit zero player gradients.

\subsection{MVP via Eig}\label{sec:bgrnd:mvp_via_eig}

Lastly, we will review how to use a series of eigendecompositions to solve an MVP: given a system of $n_e$ multivariate polynomial equations in $n_v$ variables, we would like to find their common roots. In what follows, we describe a linear algebra approach to solving the MVP based on insights from algebraic geometry~\citep{vermeersch2023block}. We assume the MVP has a finite number of affine (finite) roots, each of multiplicity $1$; we refer the reader to our references for handling more advanced cases which would relate to positive measure sets of Nash equilibria.

Let $v = [v_1, \ldots, v_{n_v}]^\top$ be a vector of the variables and $p_e(v)$, $e \in [n_e]$, denote each polynomial. A monomial in $v$ is then $v_1^{d_1} \ldots v_{n_v}^{d_{n_v}}$ with degrees given by $d_1, \ldots, d_{n_v}$. We can write any multivariate polynomial equation as a sum of monomials, each weighted by a scalar coefficient. If we choose a specific lexicographic ordering of the monomials, we can write the polynomial as $p(v) = \sum_m c_m \psi_m = c^\top \psi$ where $c$ is a vector of the coefficients and $\psi = \psi(v)$ is a vector of the ordered monomials in $v$. For example, if $v = [v_1, v_2]^\top$, then, $\psi = [1, v_1, v_2, v_1^2, v_1 v_2, v_2^2, \ldots]^\top$. The transformation from $v$ to $\psi$ is sometimes referred to as a basis expansion in the machine learning literature. It becomes clear that we can then write the entire MVP problem as $C \psi = \boldsymbol{0}$ where $C$ contains each polynomial's vector of coefficients on its rows.

If we were to compute the null space of $C$, we would find solutions that are not consistent with the desired structure of $\psi$. For example, consider the vector of monomials $[\textcolor{blue}{1}, v_1, v_2, v_1^2, \textcolor{red}{v_1v_2}, v_2^2]^\top$. $C$'s nullspace might contain a vector $[\textcolor{blue}{0}, 1, 1, 1, 1, 1]^\top$ or a vector $[1, \textcolor{codepurple}{0}, \textcolor{green}{1}, 1, \textcolor{red}{1}, 1]^\top$, but neither of these is consistent with the monomial structure we have defined because $\textcolor{blue}{1} \ne \textcolor{blue}{0}$ and $\textcolor{red}{v_1 v_2} = \textcolor{codepurple}{0} \cdot \textcolor{green}{1} \ne \textcolor{red}{1}$ respectively. A matrix of solutions $\Psi$ such that each columns obeys this monomial structure is said to be a multivariate \emph{Vandermonde} basis matrix.

First observe that if $p_e(v) = 0$, then so does $\psi_m(v) p_e(v)$ for any monomial $\psi_m(v)$. Importantly, the monomial $\psi_m(v)$ only introduces additional roots at $v = \boldsymbol{0}^\top$; zero vectors are the trivial solutions to $C\psi = \boldsymbol{0}$ and hence do not affect the multiplicity of any of the pre-existing solutions represented in the nullspace of $C$. Consider generating many more polynomial equations by this approach and considering the resulting matrix $M$, now called the Macaulay matrix (after its inventor~\citep{macaulay1902some,macaulay1916algebraic} developed it in the course of studying \emph{resultants} of polynomials). Which monomials and how many new polynomials to generate is difficult to answer, but there exists a useful upper bound that allows one to generate a sufficiently large Macaulay matrix, which naturally enforces the nullspace to contain a basis of the desired monomial structure. Let $d_i$ denote the degree of $p_e(v)$ and $d_{\max} = \max_e d_e$ the maximum degree. Define $d_{\max}^M = d_{\max} n_e - n_v + 1$ and generate all polynomials (via monomial multiplication) up to and including those of degree $d_{\max}^M$. Append these new equations to $M$. The resulting Macaulay matrix now has
\begin{align}
    n^M_r &= \sum_e \binom{d_{\max}^M - d_e + n_v}{n_v}, \quad\quad n^M_c = \binom{d_{\max} + n_v}{n_v} \label{eqn:macaulay_growth}
\end{align}
rows and columns respectively.

Now consider the vectors in the nullspace of $M$. Any linear combination of these vectors is also in the nullspace of $M$, however, a linear combination of the vectors does not necessarily match the monomial structure we are interested in recovering. For example, if $(v_1=1,v_2=1)$ and $(v_1=2,v_2=1)$ are two distinct roots of an MVP, that does not necessarily imply that, for instance, their average $(v_1=\sfrac{3}{2},v_2=1)$ is also a solution. Therefore, we want to recover the rotation of the nullspace that gives us back the precise structure we are interested in, i.e., disentangle the linear combinations into the Vandermonde basis. We can encode this requirement via a special multiplicative property of the monomials\textemdash monomials are closed under multiplication. For example, if we multiply the monomial \emph{sub}vector $[1, v_1]^\top$ by the monomial $v_1$, we generate $[v_1, v_1^2]^\top$, another monomial \emph{sub}vector of our larger $\psi$. Hence, this property can be encoded as a linear constraint:
\begin{align}
     S_{v_1} \psi &= v_1 S_1 \psi \label{eqn:gevp_1}
\end{align}
where $S_1$ is a matrix that selects out the subset of $\psi$ that represents $[1, v_1]^\top$ and similarly, $S_{v_1}$ is a matrix that selects out the subset of $\psi$ that represents $[v_1, v_1^2]^\top$.

This equation must hold for every solution $\psi$ we are interested in and for any possible monomial shift operator, not just multiplication by $v_1$, e.g., $v_l \,\, \forall l$. One might recognize~\eqref{eqn:gevp_1} fits the form of a generalized eigenvalue problem (GEVP) $Aw = \lambda Bw$: $\psi$ is a generalized eigenvector and $v_1$ its corresponding eigenvalue. Note that $v_1$ also represents a partial solution to the roots, e.g., $(v_1, v_2)$, that we are searching for, and so, eigenvalues also encode solutions (we exploit this later). We can write the generalized eigenvalue problem that solves for all generalized eigenvectors as the matrix equation:
\begin{align}
    S_{v_l} \Psi &= S_1 \Psi D_{v_l}
\end{align}
where $D_{v_l}$ contains the values of $v_l$ for each generalized eigenvector solution on its diagonal; $\Psi$ contains the generalized eigenvectors in its columns.

Let $Z$ be the nullspace of $C$. We know $\Psi$ exists within the span of this nullspace, therefore, $\Psi$ can be written as a linear combination of the columns of $Z$, i.e., $\Psi = ZW_l$ where $W_l$ is a transformation matrix. The generalized eigenvalue problem (GEVP) then becomes:
\begin{align}
    (S_{v_l} Z) W_l &= (S_1 Z) W_l D_{v_l} \label{eqn:gevp}
\end{align}
where $S_1$ is selected such that $S_1 Z$ is square and invertible. Later on, we explain how to relax this constraint without loss of generality.

Given the solution $W_l$ to the generalized eigenvalue problem in~\eqref{eqn:gevp}, we can then compute the solutions to the original MVP, $C\psi = 0$, that also obey the multiplicative shift property: $\Psi = ZW_l$. The GEVP solutions are invariant to scale, i.e., $\xi\psi$ solves the GEVP for any $\xi \in \mathbb{R}$, therefore, to recover the true corresponding monomial vector whose first entry is $v_1^0 \ldots v_{n_v}^0 = 1$, we divide the entire vector $\psi$ by its first entry. We provide pseudocode for the enitre procedure in~\Algref{alg:mvp_via_eig} assuming $0$-indexing.

\begin{algorithm}[ht]
\caption{MVP Solve via Eig Solves}
\label{alg:mvp_via_eig}
\begin{algorithmic}[1]
    \STATE Given: MVP
    \STATE Construct Macaulay matrix from MVP
    \STATE Compute null space $Z$ of Macaulay matrix \label{alg:mvp_via_eig:nullspace}
    \STATE Form $S_1$ and $Sv_l$ for some variable $v_l$ \label{alg:mvp_via_eig:shift}
    \STATE Compute eigenvectors $W_l$ of GEVP $(Sv_l Z, S_1 Z)$ \label{alg:mvp_via_eig:gevp}
    \STATE Project null space onto eigenvectors $\Psi = ZW_l$
    \STATE Columns of $\Psi$ can be normalized by first entry%
    \STATE Rows $1$ through $1 + n_v$ gives solution (each column of $\Psi$ is soln)
\end{algorithmic}
\end{algorithm}

\section{Our Formulation and Methods}

We now describe our proposed method which includes
\begin{itemize}
    \item a novel reformulation of the Nash equilibrium approximation problem as a multivariate polynomial, leading to
    \item a fast, approximate solution to Nash equilibria using least squares, and
    \item an approach to approximately recovering all Nash equilibria via a series of stochastic singular value decompositions and power iteration.
\end{itemize}

\subsection{NE via MVP}\label{sec:ne_mvp}

As explained in Section~\ref{sec:bgrnd:ne_tsallis}, games with utilities that are regularized with sufficient Tsallis entropy bonuses yield equilibria with full support. By properties of vector norms, $\gamma = ||\nabla^i_{x_i}||_{1/\tau} \le ||\nabla^i_{x_i}||_1 \le \vert \mathcal{A}_i \vert$ (recall Assumption~\ref{assump:u_01}). Intuitively, by replacing $\gamma$ with this upper bound, we increase the strength of the Tsallis regularization, further pulling all NE closer to the uniform distribution. This step is critical as it also enables us to replace the complex (non-polynomial) norm with a constant (i.e., a polynomial of degree $0$), thereby ensuring player utilities and gradients are multivariate polynomials in the strategies $x$.

We now state a result that shows that for sufficiently \emph{low} parameter $\tau$, approximate equilibria in the Tsallis-transformed game achieve low exploitability ($\epsilon$) in the original game:
\begin{align}
    \epsilon_i(\boldsymbol{x}) &= \max(\nabla^i_{x_i}) - u_i(\boldsymbol{x})
    \\ &\le H^{\tau}(x_i) + \tau H^{\tau}(\br^{\tau}_i) + \sqrt{2} ||\Pi_{T\Delta}(\nabla^{i\tau}_{x_i})||_2
    \\ &\le \tau \vert \mathcal{A}_i \vert \ln(\vert \mathcal{A}_i \vert) + \sqrt{2} ||\Pi_{T\Delta}(\nabla^{i\tau}_{x_i})||_2
\end{align}
and $\epsilon = \max_i \epsilon_i \le \sum_i \epsilon_i$ completing the bound.

Note that $||\Pi_{T\Delta}(\nabla^{i\tau}_{x_i})||_2 = 0$ for an exact NE in the Tsallis-transformed game. This means that solving for the Nash equilibrium of the game with $H^{\tau}$ entropy bonuses well approximates the original Nash equilibrium problem at low $\tau$. Constructing player $i$'s zero-gradient condition from equations~\plaineqrefp{eqn:grad} and~\plaineqrefp{eqn:projgrad}, we find
\begin{align}
    \Pi_{T\Delta}\big( \nabla^{i\tau}_{x_i} \big) &= [I - \frac{1}{\vert \mathcal{A}_i \vert} \mathbf{1} \mathbf{1}^\top] (\nabla^i_{x_i} - \vert \mathcal{A}_i \vert x_i^{\tau}) = 0.
\end{align}

Now perform a variable substitution; assume $1/\tau$ is an integer and let $v_i = x_i^{\tau}$. Then $x_i = v_i^{1/{\tau}}$ and for all $i$
\begin{align}
    \Pi_{T\Delta}\big( \nabla^{i\tau}_{x_i} \big) &= [I - \frac{1}{\vert \mathcal{A}_i \vert} \mathbf{1} \mathbf{1}^\top] (U^{(i)}_{1 \ldots N} \ldots v_{i-1}^{1/\tau} v_{i+1}^{1/\tau} \ldots - \vert \mathcal{A}_i \vert v_i) = 0 \label{eqn:zero_proj_grad_poly}
\end{align}
where we have made use of the Einstein notation introduced in Section~\ref{sec:bgrnd},~\eqref{eqn:gradi_einstein}. Going forward, for clarity, we will alternate between subscripting variables according to their player strategy correspondence $v_{i, a_i}$ and their flattened representation $v_l$; their meaning should be clear from the context.

First recognize that by projecting player $i$'s gradient, i.e., subtracting the mean using the rank-($\vert \mathcal{A}_i \vert - 1$) matrix $[I - \frac{1}{\vert \mathcal{A}_i \vert} \mathbf{1} \mathbf{1}^\top]$, we have effectively reduced the number of linear independent equations by $1$, and so we can arbitrarily drop one of the zero-gradient equations, leaving us with $\vert \mathcal{A}_i \vert - 1$ equations for each player. Next, notice that each equation is a polynomial of degree $(n-1)(1/\tau)$.

In addition to the zero-gradient conditions, we can include the simplex constraints:
\begin{align}
    \sum_{a_i \in \mathcal{A}_i} v_{i,a_i}^{1/\tau} &= 1 \label{eqn:sum_to_1}
    \\ v_{i,a_i} &\ge 0 \,\, \forall \,\, a_i \in \mathcal{A}_i.
\end{align}

The entire set of equations (\ref{eqn:zero_proj_grad_poly}-\ref{eqn:sum_to_1}) represents a multivariate system of $\sum_i \vert \mathcal{A}_i \vert$, degree-$(n-1)(1/\tau)$ polynomial equations in $\sum_i \vert \mathcal{A}_i \vert$ unknowns whose real, nonegative\footnote{If we want to remove the non-negativity constraints, we can replace $v_i$ with $z_i^2$.} solutions well approximate the Nash equilibria of the original game.

\paragraph{Other MVP representations of NE} Prior formulations~\citep{sturmfels2002solving} require solving an MVP of degree $n$ subject to nonlinear constraints of degree $n-1$. This approach is inefficient if a large portion of the candidate roots returned by an MVP algorithm violate these constraints and must be discarded. Our approach removes all complex constraints, leaving a pure MVP.
This means one can simply search for roots in the real, positive orthant (an exponential reduction in the search space).

In addition, past PCP formulations result in MVPs with degree at least $n$, whereas ours results in an MVP with degree $(n - 1) \tau^{-1}$ with $\tau^{-1} \in \mathbb{N}$. For $\tau=1$, this results in a linear system for all $2$-player games. To our knowledge, our formulation implies the first least-squares approach to approximating NE in general-sum games with approximation guarantees. We examine the performance of this approach later in Section~\ref{sec:experiments}.

\subsection{MVP via SVD}
We now describe how to transform the procedure detailed in Section~\ref{sec:bgrnd} into one that relies only on calls to top-$k$ singular value decomposition (SVD) solvers or power iteration, and in particular, iterative stochastic variants of these solvers. There are two steps that appear in~\Algref{alg:mvp_via_eig} requiring eigenvalue decompositions. We specifically target SVD and power iteration as our core primitive routines as these have drawn enormous attention within both numerical linear algebra and machine learning, and we hope techniques and insights developed there can shed new light on this routine that is traditionally inspired by algebraic geometry. We make the following assumption which translates the assumption of distinct, affine roots from Section~\ref{sec:bgrnd:mvp_via_eig} into the parlance of equilibria.
\begin{assumption}[Isolated NEs]\label{assump:isolated_nes}
All NEs are isolated. Also for any two NEs $x^{(s)}$ and $x^{(s')}$, $x^{s}_{i,a_i} \ne x^{s'}_{i,a_i} \,\, \forall i, a_i \in \mathcal{A}_i$.
\end{assumption}
Assumption~\ref{assump:isolated_nes} can be overcome if we include a more general shift matrix $S_{g}$ where $g$ is any polynomial in $x$ and repeat line~\ref{alg:mvp_via_eig:shift} onward for each shift operator $v_l$ and $g$~\citep[Sections 2.4.2-2.4.3]{vermeersch2023block}.

First, line~\ref{alg:mvp_via_eig:nullspace} of~\Algref{alg:mvp_via_eig} computes the null space $Z$ of the Macaulay matrix $M$. Let $\lambda_M^*$ be any value greater than the squared maximum singular value of $M$ (equiv. the maximum eigenvalue of $M^\top M$)\footnote{The SVD of a matrix $A$ is related to the eigendecomposition of the symmetric, positive semi-definite matrix $A^\top A$ in that the right singular vectors of the former match the eigenvectors of the latter, and the \emph{squared} singular values match the eigenvalues.}. Also let the dimensionality of the null space of $M$ be $d_M$. Then approximating the nullspace of $M$ can be recast as computing the top-$d_M$ singular vectors of $\tilde{M} = \lambda_M^* I - M^\top M$. By inspection, if $z$ is a unit-vector in the null space of $M$, then $z^\top \tilde{M} z = \lambda_M^*$ which is necessarily the maximum eigenvalue of $\tilde{M}$ because $z^\top M^\top M z \ge 0$. Furthermore, we need not assume knowledge of $\lambda_M^*$ or $d_M$. The maximum singular value of $M$ can be computed via a top-$1$ SVD solver. The integer $d_M$ can be discovered by running the top-$k$ solver on $\tilde{M}$ with successively larger values of $k$, e.g., doubling $k$. Once the solver returns singular vectors with corresponding nonzero singular values, we can halt the process and return only the vectors with zero singular value\textemdash these vectors constitute the null space of $M$.

\begin{algorithm}[ht]
\caption{Approximate NE via SVD (and a little power iteration)}
\label{alg:ne_via_svd}
\begin{algorithmic}[1]
    \STATE Given: Normal-form game $\mathcal{G}$
    \STATE Construct MVP from $\mathcal{G}$ via equations~(\ref{eqn:zero_proj_grad_poly}-\ref{eqn:sum_to_1})
    \STATE Construct Macaulay matrix from MVP
    \STATE Compute null space $Z$ of Macaulay matrix using SVD
    \STATE Form $S_1$ and $Sv_l$ for some $v_l$ such that $\texttt{rank}(S_1 Z) = \texttt{dim}(Z)$
    \STATE Compute $Q=(S_1 Z)^{\dagger}$ using SVD
    \STATE Solution set $\mathcal{X}^* = \{ \}$
    \FOR{$x \in \{0, \ldots, 1\}$}
        \STATE $\lambda = x^{\tau}$
        \STATE Compute intermediate matrix $R_{\lambda} = Q Sv_l Z - \lambda I$
        \STATE Compute $R_{\lambda}^{\dagger}$ using SVD
        \STATE Compute largest eigenvector $w$ of $R_{\lambda}^{\dagger}$ using power iteration
        \STATE Project null space onto eigenvector $\psi = Zw$
        \STATE Normalize $\psi$ by first entry, e.g., $\psi /= \psi[0]$
        \STATE Collect rows $1$ through $1 + \sum_i \vert \mathcal{A}_i \vert$ for candidate solution
        \STATE Let $x = \psi[1:1+\sum_i \vert \mathcal{A}_i \vert]^{1/\tau}$
        \IF{$x \ge 0$ and $x$ is new solution and power iteration converged}
            \STATE Add $x$ to solution set, $\mathcal{X}^* = \mathcal{X}^* \cup \{x\}$
        \ENDIF
    \ENDFOR
    \STATE Return $\mathcal{X}^*$
\end{algorithmic}
\end{algorithm}

The second call to a GEVP in line~\ref{alg:mvp_via_eig:gevp} of~\Algref{alg:mvp_via_eig} requires a more involved translation. Note that technically,~\eqref{eqn:gevp} only conforms to the definition of a generalized eigenvalue problem if both $S_{v_l} Z$ and $S_1 Z$ are square. However, this is not strictly necessary for~\eqref{eqn:gevp} to have solutions. It is possible to overpopulate the number of equations in~\plaineqrefp{eqn:gevp} such that the rows exceed the columns of $Z$. As long as the number of linearly independent rows of $S_1 Z$ equals the number of columns of $Z$, we will retain the same solution set. We state this as it can be easier in practice to select rows of $Z$ less judiciously and instead transform the problem into the following using the pseudoinverse $\dagger$:
\begin{align}
    (S_1 Z)^\dagger (S_{v_l} Z) W_l &= W_l D_{x_1}. \label{eqn:asymeig}
\end{align}

Hence, we first compute the pseudoinverse of $S_1 Z$. The pseudoinverse of a matrix $A$ can be computed $(A^\top A)^{-1} A^T$. The inverse of the symmetric matrix $A^\top A$ can be computed via its eigendecomposition, or equivalently via the SVD of $A$. Let $\Sigma$ contain the singular values of $A$ and $W$ its right singular vectors. Then $(A^\top A)^{-1} = W \Sigma^{-2} W^\top$. If an approximate inverse is tolerable, then only the bottom few singular values and vectors of $A$ need be computed, and the matrix inverse can be similarly reconstructed from them. The result is that we can now transform our challenge into computing the eigenvectors of $(S_1 Z)^\dagger (S_{v_l} Z)$, which is of shape $d_M \times d_M$.

To compute the eigenvectors of this \emph{asymmetric} matrix, we now employ a technique called \emph{inverse iteration}. Assume $\mu$ well approximates a true eigenvalue $\mu^*$ of a matrix $A$. Then the matrix $(A - \mu I)^{-1}$ will achieve its maximum eigenvalue at a vector $w$ which is also the eigenvector of $A$ corresponding to the true $\mu^*$. The pseudoinverse equals the inverse if $(A - \mu I)^{-1}$ is invertible; therefore, we can first compute $(A - \mu I)^{\dagger}$ using SVD as explained before. Given the pseudoinverse, we can then run power iteration (initialized with a random \emph{real} vector) to approximate the eigenvector $w \in \mathbb{R}^{d_M}$. This eigenvector $w$ represents one solution (column) of $W$. As one would suspect, this approach seems most useful when a good approximation of $\mu^*$ is already known. As we now explain, we can greatly reduce the search space over possible $\mu$ for the Nash equilibrium problem we are interested in.

\begin{figure*}[ht!]
    \centering
    \includegraphics[width=\textwidth]{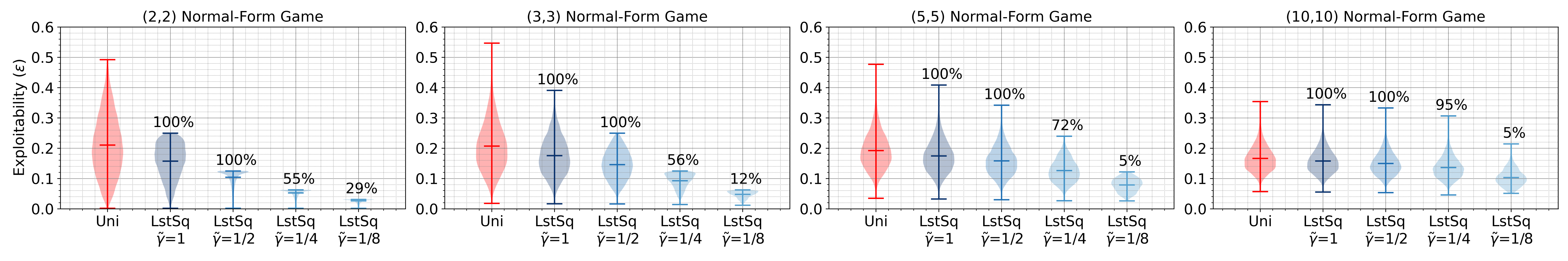}
    \caption{(Least Squares vs Uniform) Violin plots of the exploitability attained by our proposed least-squares method (LstSq) versus a uniform strategy profile (Uni) on $10$ thousand games of each size with payoffs normalized to $[0.001, 1]$. Parameter $\tilde{\gamma} = \gamma / \vert \mathcal{A}_i \vert$. Percentages indicate the success rate of LstSq returning a valid output that lies on the product simplex.}
    \label{fig:least_squares}
\end{figure*}

Recall that the eigenvalues $D_{v_l}$ of equations~\plaineqrefp{eqn:gevp} and~\plaineqrefp{eqn:asymeig} also represent partial solutions to the roots we are computing. In Section~\ref{sec:ne_mvp}, we transformed the probabilities under strategies in the equilibrium into new variables with $v_l = x_l^{\tau}$. Each entry in $x_{l}$ lies in $[0, 1]$ and $\tau \ge 0$, therefore, each entry in $v_l$ lies in $[0, 1]$ as well. So we know that all $\mu^*$ lie in $[0, 1]$ in our case. A practical approach is then to search over the unit interval, employing inverse iteration with each value of $\mu$. If power iteration converges to an eigenvector $w \in \mathbb{R}^{d_M}$, then we can examine the $n_v$ entries of $x = (Zw)^{\sfrac{1}{\tau}}$ corresponding to the degree-$1$ monomials to determine a) do they constitute an equilibrium profile (i.e., do they lie on the product simplex) and b) are they distinct from previous solutions we have already recovered. We summarize the procedure described above in~\Algref{alg:ne_via_svd} assuming $0$-indexing.

\section{Experiments}\label{sec:experiments}

We evaluate our proposed approaches in two different settings. In the first, we test our least-squares method on randomly sampled $2$-player, general-sum games. In the second, we test the ability of stochastic singular value decomposition and power iteration to recover all equilibria of the classic Chicken game. To our knowledge, this is the first work investigating solving MVPs via \emph{stochastic} eigendecompositions, let alone solving NEPs in this way.

\subsection{$2$-player NE via Least Squares}

We compare the exploitability of the approximate equilibrium returned by our least squares approach to that of the uniform strategy profile on $2$-player games with $2$, $3$, $5$, and $10$ actions per player. For each game size, we sample a payoff matrix for each player with entries in $[0, 1]$. We then normalize all payoffs such that the minimum and maximum payoff (across both players) is $0.001$ and $1$ respectively. This is to ensure Assumption~\ref{assump:u_01} is met. We repeat this process ten thousand times for each game size and report the distribution of exploitability values observed in Figure~\ref{fig:least_squares}.

For small games, our least squares method improves beyond the uniform strategy baseline, however, for larger action spaces, this gap decays. Our theory guarantees our least-squares method actually returns valid mixed strategies for $\gamma=\vert \mathcal{A}_i \vert = 2$. Define the normalized parameter $\tilde{\gamma} = \gamma / \mathcal{A}_i \vert$. For $\tilde{\gamma} < 1$, it is possible the returned strategy profile lies off the simplex for at least one of the players. We find that the method is able to return improved approximations to NEs for lower values of $\tilde{\gamma}$ in some cases without failure. For sufficiently low $\tilde{\gamma}$, the success rate drops significantly. Furthermore, similar to $\tilde{\gamma}=1$, the improvement is tempered for larger action spaces.

\begin{table}[ht!]
    \centering
    \begin{tabular}{l||c|c|c|c}
        Batch Size & \% Macaulay & \% $S_1 Z$ & Success & Jensen-Shannon \\ \hline\hline
        1000 & 100\% & 100\% & 0.90 & 0.013 \\
        500 & 59\% & 100\% & 0.90 & 0.013 \\
        400 & 48\% & 81\% & 0.81 & 0.016 \\
        200 & 24\% & 40\% & 0.79 & 0.016 \\
        100 & 12\% & 20\% & 0.91 & 0.014
    \end{tabular}
    \caption{(Stochastic Eigendecomposition of Chicken with Varying Batch Size) We report the average performance of our approach in approximately recovering the three NEs of the Tsallis-transformed Chicken game. The first column reports the batch size used in the stochastic SVD and power iteration steps. The second and third translate the batch size into a percentage of the number of rows in the Macaulay ($M$) and $S_1 Z$ matrices respectively ($R_{\lambda}$ contains only $d_M=81$ rows which is below our lowest batch size). The Success column reports the success rate of the algorithm returning $3$ distinct NEs. The last column reports the average Jensen-Shannon distance of the returned strategy profiles to the ground truth NEs of the Tsallis-transformed Chicken game. Averages are computed over $100$ trials. The \# of iterations was increased for batch sizes below $400$.}
    \label{tab:chicken_eigengame}
\end{table}

\begin{table}[ht!]
    \vspace{-0.5cm}
    \centering
    \begin{tabular}{l||c|c|c|c}
        \multirow{2}{*}{$\mathcal{X}^*$} & \multicolumn{2}{c|}{Player 1} & \multicolumn{2}{c}{Player 2} \\
         & $x_{11}$ & $x_{12}$ & $x_{21}$ & $x_{22}$ \\ \hline \hline
        $x^{(1)*}$ & $0.995$ & $0.005$ & $0.107$ & $0.893$ \\
        $x^{(2)*}$ & $0.107$ & $0.893$ & $0.995$ & $0.005$ \\
        $x^{(2)*}$ & $0.597$ & $0.403$ & $0.597$ & $0.403$ \\
    \end{tabular}
    \caption{(Ground Truth NEs of Chicken) We use~\Algref{alg:mvp_via_eig} to uncover the true three approximate NEs for Chicken.}
    \label{tab:chicken_gt}
\end{table}

\subsection{NE via Stochastic Eigendecomposition}

Next, we evaluate~\Algref{alg:ne_via_svd} on a Chicken game:
\begin{align}
U^{(1)} &= \begin{bmatrix}
0.7527 & 0.505 \\
1.0 & 0.01
\end{bmatrix}
&U^{(2)} = \begin{bmatrix}
0.7527 & 1.0 \\
0.505 & 0.01
\end{bmatrix}
\end{align}
where the first row / column represents the ``swerve'' action and the second the ``go straight'' action.

\begin{figure}[ht]
    \centering
    \includegraphics[width=0.4\textwidth]{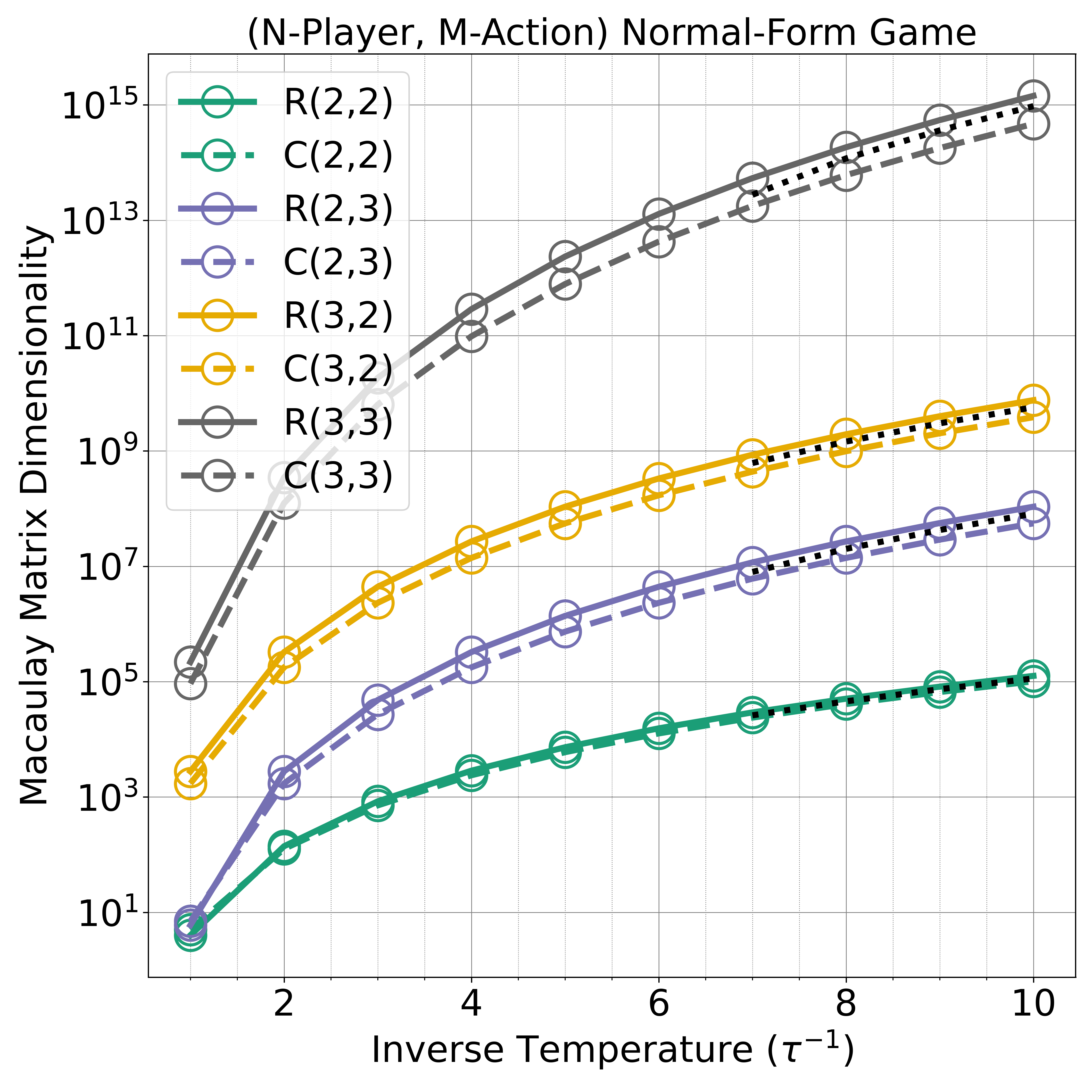}
    \caption{(Macaulay Matrix Growth) An upper bound on the number of rows R and columns C is shown for $2$-player games with various action spaces ($\vert \mathcal{A}_1 \vert, \vert \mathcal{A}_2 \vert$) as inverse temperature $\tau^{-1}$ is varied. The Macaulay matrix can quickly grow to sizes beyond typical memory limits (see~\eqref{eqn:macaulay_growth} for formulae), however, for fixed game sizes, this growth is actually polynomial for large $\tau^{-1}$, not exponential. The best fit dotted lines provide a visual aid\textemdash R and C behave as $\tilde{\mathcal{O}}(\tau^{-\sum_i \vert \mathcal{A}_i \vert})$.}
    \label{fig:macaulay_growth}
\end{figure}

Motivated by the least-squares results, we set $\tilde{\gamma}=\sfrac{1}{4}$ despite the lack of theoretical guarantee that interior NEs will exist. We found this gave better performance similar to what we observed in Figure~\ref{fig:least_squares}. For our stochastic SVD routine, we used the recently developed EigenGame Unloaded method~\citep{gemp2021eigengame}\textemdash we discuss this particular choice in the next section and report hyperparameter settings in~\appref{app:hyper}. We compute ground truth using~\Algref{alg:mvp_via_eig} with eigendecompositions implemented in Scipy~\citep{2020SciPy-NMeth}.

Table~\ref{tab:chicken_eigengame} shows that while the success rate even in the full batch regime is not $100\%$, the method still succeeds for at least $79\%$ of the trials even for quite low batch size percentages. This result should not be taken for granted. The finding that null spaces and pseudo-inverses can be computed approximately and yet still retain information about the location of Nash equilibria is important. A level of robustness is critical for one to explore a variety of scalable, stochastic techniques.

\section{Discussion}\label{sec:disc}

We now discuss limitations of our proposed implementation, how to complete the itinerant loop illustrated in Figure~\ref{fig:itinerant_loop}, as well as implications for biological plausibility of Nash equilibrium approximation in the brain.

\subsection{Limitations}\label{sec:disc:lim}

Figure~\ref{fig:macaulay_growth} shows that the Macaulay matrix $M$ we construct to solve for NEs quickly grows to a size beyond standard hardware limits. For fixed a game size (i.e., fixed $n_v = \sum_i \vert \mathcal{A}_i \vert$), the binomial coefficients in~\eqref{eqn:macaulay_growth} governing the size of $M$ grow polynomially with degree $n_v$ in their numerator (``selector'') which is linear in $\tau^{-1}$; specifically, the rows and columns grow as $\tilde{\mathcal{O}}(\tau^{-\sum_i \vert \mathcal{A}_i \vert})$ hiding lower order terms. Note that Algorithm~\ref{alg:mvp_via_eig} computes an eigendecomposition of this matrix, which has a complexity that is cubic in its size~\citep{pan1999complexity}. Hence the complexity of this step is $\tilde{\mathcal{O}}(\tau^{-3\sum_i \vert \mathcal{A}_i \vert})$, which is polynomial in $\tau^{-1}$ for a fixed game size but exponential in the game size.

Despite the fast growth of $M$, the number of nonzeros in each row of $M$ is small meaning the effective column size is low. Recall that the initial seed polynomials $p_e(x)$ defined in~\eqref{eqn:zero_proj_grad_poly} indicate the number of nonzero coefficients in each row should be at most $\Pi_{i=1}^N \vert \mathcal{A}_i \vert$, i.e., the size of a player's payoff tensor. This property is interesting because machine learning research has specifically developed solvers with complexity that depends on the number of nonzeros in a matrix $A$ rather than the dense size~\citep{allen2016lazysvd}, e.g., $\mathcal{O}(nnz(A) + poly(1/\delta))$ where $\delta$ is the desired error rate. On the other hand, the number of rows in $M$ is still $\mathcal{O}(\tau^{-\sum_i \vert \mathcal{A}_i \vert})$ but again, is handled by the standard stochastic ML model where batches of rows are sampled at random and fed to an iterative update rule. The implication is that the null space of $M$ can be approximated efficiently and with low memory. As long as the null space of $M$ is small (for reference the null space of $M$ for Chicken is spanned by $d_M = 81$ vectors), the remaining eigendecompositions become manageable (e.g., matrices have $d_M = 81$ columns). We leave exploration of these improvements to future work, as we have chosen to use the approach developed in EigenGame for two reasons we discuss next.

\subsection{Completing the Loop}\label{sec:disc:loop}

Prior work~\citep{gempeigengame,gemp2021eigengame} showed top-$k$ SVD of a matrix $A \in \mathbb{R}^{d_r \times d_c}$ can be formulated as an \emph{EigenGame} where $k$ players each select a unit-vector in $\mathbb{R}^{d_c}$ to maximize a utility function that rewards vector directions $w$ with high Rayleigh quotients ($w^\top A w$) and penalizes for alignment to other vectors ($\langle w_i, A w_j\rangle^2 / \langle w_j, A w_j\rangle$ for all $i \ne j$). Modified versions of this game are \emph{iterated dominance solvable} and iterative solvers have been designed that converge to the top-$k$ eigenvectors, the strict, pure Nash equilibrium of the game when $A$ contains distinct eigenvalues. The strategy space for each player is the sphere $\mathbb{S}^{d_c-1}$, a continuous Riemannian manifold, and so is not a \emph{finite} game.

\begin{figure}
    \centering
    \includegraphics[width=0.35\textwidth]{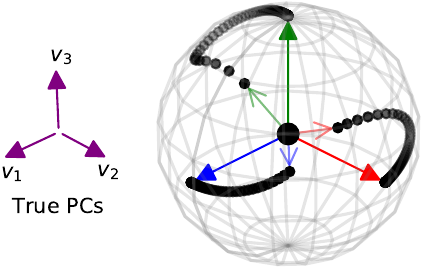}
    \caption{(EigenGame\textemdash Figure 2 reproduced from~\citep{gempeigengame} with permission) Players select unit vectors from a continuous set (the sphere). Trajectories show the dynamics of players updating their strategies simultaneously with gradient descent. A wireframe mesh can be overlayed on the sphere to convert it into a game with a finite strategy set~\citep{assos2023online}.}
    \label{fig:eigengame}
\end{figure}

Nevertheless, one can construct an approximation of this game using techniques developed in empirical game-theoretic analysis (EGTA)~\citep{warsh2002analyzing,wellman2006methods} similar to techniques pervasive across engineering disciplines such as finite element methods. Some of the grandest achievements in games~\citep{vinyals2019alphastar,silver2016mastering} leveraged this viewpoint and a stylized version of their approach~\citep{lanctot2017unified} has been proven to converge to approximate equilibria in infinite games~\citep{assos2023online}. The idea is to select a finite set of strategies from the continuous strategy set to construct a normal-form game; this set can form a mesh over each sphere, for example (see Figure~\ref{fig:eigengame}). As this mesh is refined, the approximation to the continuous setting becomes more accurate. It is in this sense that we can recover a finite normal form game that approximates the SVD problem.

\begin{figure}
    \centering
    \includegraphics[width=0.4\textwidth]{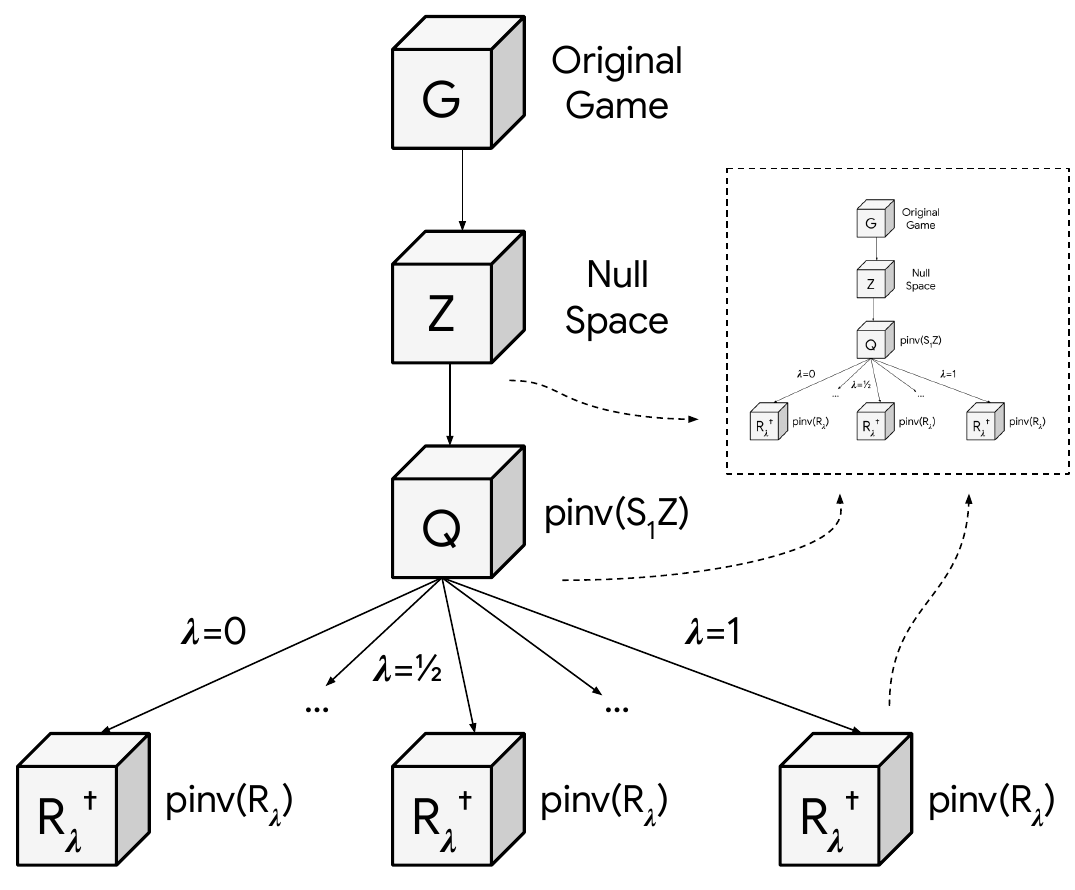}
    \caption{(A Tree of Games) Every normal-form Nash equilibrium problem $G$ can be decomposed into a tree of simpler (iterative dominance solvable) normal-form games albeit of higher dimensionality. This decomposition can be continued recursively. Each cube is a normal-form game NEP.}
    \label{fig:tree_of_games}
\end{figure}

The implication is that every normal form game can be decomposed recursively into a fractal tree of games illustrated in Figure~\ref{fig:tree_of_games}. These derived games are simpler\textemdash they are iterative dominance solvable\textemdash but they are of increasing dimensionality and hence possibly impractically suited to a divide-and-conquer approach.

\balance{}

\subsection{Biological Plausibility}\label{sec:disc:bio}

Another reason for selecting EigenGame Unloaded~\citep{gemp2021eigengame} as our SVD solver is that it has been connected to the Generalized Hebbian Algorithm (GHA)~\citep{sanger1989optimal,gang2019fast,chen2019constrained}. Hebbian theory is a theory of synaptic plasticity in the brain and is often summarized by the mnemonic ``neurons that fire together, wire together''~\citep{hebb2005organization,lowel1992selection}. GHA is a neurobiologically plausible Hebbian learning rule~\citep{lillicrap2020backpropagation} that has been proposed as a method by which connectionist systems might solve a wide array of learning tasks including dimensionality reduction, correlation analysis, feature discovery, linear problem solving, and other fundamental statistical challenges.

To our knowledge, this is the first neurobiologically plausible learning rule for approximating Nash equilibria in general-sum games. The most closely related research on replicator dynamics is studied in the context of evolutionary game theory~\citep{gintis2000game,weibull1997evolutionary} as plausible dynamics that populations may follow, however, these do not necessarily approximate Nash equilibria beyond the $2$-player, zero-sum setting (although their stationary distributions are of separate interest~\citep{omidshafiei2019alpha}) nor have they been matched to theories of neural plasticity.

\subsection{MVP Perspective \& Algebraic Geometry}\label{sec:disc:alg}
Most methods for approximating equilibria focus on guaranteeing convergence to just one NE. But, as ~\citet{sturmfels2002solving}, algebraic tools developed to compute NEs treat the set of NEs holistically, returning all roots at once. Furthermore, they can provide bounds on the precise number of equilibria. For example, the MVP structure for an $(N\ge2)$-player, $2$-action game limits the number of isolated, interior NEs to exactly  $1, 2, 9, 44, 265, 1854, 14833, 133496, \ldots$. We hope our novel MVP formulation can uncover similar insights.

\subsection{Future Work}

In future work, it seems sensible to explore least squares as a useful initialization for warm starting other NE algorithms. We also want to understand how approximation error propagates through the sequence of eigenvalue problems in~\Twoalgref{alg:mvp_via_eig}{alg:ne_via_svd}.

As mentioned in the discussion, we expect the pursuit of null space methods from machine learning with polynomial dependence on the number of nonzeros to be fruitful. And although~\Algref{alg:ne_via_svd} replaces all eigendecompositions with iterative SVD solvers, it might be the case that the most practically efficient approach is to only use iterative methods to compute the null space $Z$, and then relying on exact linear algebra routines thereafter.

Aside from efficiency, SVD has been extended to nonlinear settings with either kernel-tricks~\citep{scholkopf1998nonlinear} and or deep learning (autoencoders)~\citep{kramer1991nonlinear,kingma2013auto}. Can we learn useful deep representations of games in this way? In a separate vein, can we extend this approach to games with continuous strategy spaces by using deep learning approaches to learn eigenfunctions of infinite matrices~\citep{pmlr-v162-deng22b,pfauspectral}?

\section{Conclusion}

This paper identifies two novel, interdisciplinary approaches to approximating Nash equilibria in games. The first reduces NE approximation in $2$-player, general-sum games to a least squares problem, that can return a solution quickly. The second prescribes an iterative, stochastic approach that solves NEs as a multivariate polynomial problem through a series of eigendecompositions. We discussed limitations of this approach as well as promising avenues for future research. Lastly, we opened up a discussion on biological plausibility of our proposed method, which hinges on connections to Hebbian learning. We see a primary contribution of this work as collating the knowledge necessary to understanding this style of method in the context of game theory as well as doing the groundwork to enable future extensions of this new family of approaches.

\bibliographystyle{ACM-Reference-Format} 
\bibliography{bib}

\newpage
\onecolumn
\appendix
\section{Nash of a Tsallis Regularized Game as a System of Multivariate Polynomials}\label{appx}

\citet{gemp2022sample} assume all payoffs are in $(0,1]$ and consider $H^{\tau}(x_i) = \frac{\gamma}{\tau+1}(1 - \sum_l x_{il}^{\tau+1})$ with $\tau \in [0,1]$. Recall equilibria are guaranteed to exist in $n$-player concave-utility games~\citep{rosen1965existence}. The best response $\br_i^{\tau}$ under utilities regularized with $H^{\tau}$ with $\gamma = \big(\sum_l (\nabla^i_{il})^{1/\tau}\big)^{\tau} = ||\nabla^i_{x_i}||_{1/\tau}$ can be derived by setting the gradient to zero and solving for $x_i$:
\begin{align}
    \nabla_{x_i} u_i^{\tau}(\boldsymbol{x}) &= \nabla^i_{x_i} - \gamma x_i^{\tau} = 0 \implies \br^{\tau}_i = \Big( \frac{\nabla^i_{x_i}}{||\nabla^i_{x_i}||_{1/\tau}} \Big)^{1/\tau} = \frac{(\nabla^i_{x_i})^{1/\tau}}{\sum_l (\nabla^i_{il})^{1/\tau}} \in \Delta^{\vert \mathcal{A}_i \vert-1}.
\end{align}

Now, we can infer that if we increase the strength of the regularization used in~\citep{gemp2022sample}, the best response will continue to remain in the interior. Whereas if we reduced the strength, it's possible the best response would then lie on the boundary (and would evade a closed-form solution). Note that $\gamma \le \vert \mathcal{A}_i \vert$ for payoffs in $[0, 1]$ and $\tau \in (0, 1]$ because $||z||_{1/\tau} \le ||z||_{1}$ for any $z$ and $\tau \in (0, 1]$ and particular $z=\nabla^i_{x_i}$. Therefore, if set $\gamma=\vert \mathcal{A}_i \vert$, we know that the best response will remain in the interior. However, the gradient at the best response may not necessarily be $0$. In general, gradients at interior best responses satisfy a more general property:
\begin{align}
    \nabla_{x_i} u_i^{\tau}(\boldsymbol{x}) &= \nabla^i_{x_i} - \vert \mathcal{A}_i \vert x_i^{\tau} = c \mathbf{1}_i \label{eqn:abr_prop}
\end{align}
where $c \in \mathbb{R}$ is some scalar. This implies there exists a $c \in \mathbb{R}$ such that
\begin{align}
    \br^{\tau}_i &= \Big( \frac{\nabla^i_{x_i} - c \mathbf{1}_i}{\vert \mathcal{A}_i \vert} \Big)^{1/\tau}. \label{eqn:abr}
\end{align}

For now, assume $1/\tau$ is an odd integer. This implies the numerator in~\eqref{eqn:abr} must be nonnegative for the result to lie on the simplex. Therefore, we can further restrict $c \le \min(\nabla^i_{x_i})$. Likewise, the numerator must not be greater than $\vert \mathcal{A}_i \vert$. Therefore, $c \ge \max(\nabla^i_{x_i}) - \vert \mathcal{A}_i \vert$. Together, we find $c \in [\max(\nabla^i_{x_i}) - \vert \mathcal{A}_i \vert, \min(\nabla^i_{x_i}) ]$.

Now let's consider the difference in payoff achieved at an un-regularized best response with that of a regularized best response. We would like to obtain a result
\begin{align}
    \max (\nabla^i_{x_i}) - \langle \nabla^i_{x_i}, \argmax_{x_i} \{ x_i^\top \nabla^i_{x_i} + H^{\tau}(x_i) \} \rangle &\le a \tau^\xi
\end{align}
for some $a > 0$ and integer exponent $\xi$ meaning that a regularized best response achieves approximately the same utility as a true best response as $\tau \rightarrow 0^+$. Plugging~\eqref{eqn:abr} into the regularized utility function gives
\begin{align}
    u_i^{\tau}(\br^{\tau}_i, x_{-i}) &= \langle \Big( \frac{\nabla^i_{x_i} - c \mathbf{1}_i}{\vert \mathcal{A}_i \vert} \Big)^{1/\tau} , \nabla^i_{x_i} \rangle + \frac{\vert \mathcal{A}_i \vert}{\tau+1}(1 - \sum_l \Big( \frac{\nabla^i_{x_{il}} - c}{\vert \mathcal{A}_i \vert} \Big)^{\frac{\tau+1}{\tau}})
    \\ &= \sum_l \Big( \frac{\nabla^i_{x_{il}} - c}{\vert \mathcal{A}_i \vert} \Big)^{1/\tau} \nabla^i_{x_{il}} + \frac{\vert \mathcal{A}_i \vert}{\tau+1}(1 - \sum_l \Big( \frac{\nabla^i_{x_{il}} - c}{\vert \mathcal{A}_i \vert} \Big)^{\frac{\tau+1}{\tau}})
    \\ &= \frac{\vert \mathcal{A}_i \vert}{\tau+1} + \sum_l \Big( \frac{\nabla^i_{x_{il}} - c}{\vert \mathcal{A}_i \vert} \Big)^{1/\tau} \nabla^i_{x_{il}} - \frac{\vert \mathcal{A}_i \vert}{\tau+1} \Big( \frac{\nabla^i_{x_{il}} - c}{\vert \mathcal{A}_i \vert} \Big)^{\frac{\tau+1}{\tau}})
    \\ &= \frac{\vert \mathcal{A}_i \vert}{\tau+1} + \sum_l \Big( \frac{\nabla^i_{x_{il}} - c}{\vert \mathcal{A}_i \vert} \Big)^{1/\tau} \Big( \nabla^i_{x_{il}} - \frac{\vert \mathcal{A}_i \vert}{\tau+1} \frac{\nabla^i_{x_{il}} - c}{\vert \mathcal{A}_i \vert} \Big)
    \\ &= \frac{\vert \mathcal{A}_i \vert}{\tau+1} + \sum_l \Big( \frac{\nabla^i_{x_{il}} - c}{\vert \mathcal{A}_i \vert} \Big)^{1/\tau} \Big( \nabla^i_{x_{il}} - \frac{\nabla^i_{x_{il}} - c}{\tau+1} \Big)
    \\ &= \frac{\vert \mathcal{A}_i \vert}{\tau+1} + \sum_l \Big( \frac{\nabla^i_{x_{il}} - c}{\vert \mathcal{A}_i \vert} \Big)^{1/\tau} \Big( \frac{(\tau+1) \nabla^i_{x_{il}} - \nabla^i_{x_{il}} + c}{\tau+1} \Big)
    \\ &= \frac{\vert \mathcal{A}_i \vert}{\tau+1} + \sum_l \Big( \frac{\nabla^i_{x_{il}} - c}{\vert \mathcal{A}_i \vert} \Big)^{1/\tau} \Big( \frac{\tau\nabla^i_{x_{il}} + c}{\tau+1} \Big)
\end{align}

We can lower bound this utility using the lower bound for $c$:
\begin{align}
    u_i^{\tau}(\br^{\tau}_i, x_{-i}) &\ge \frac{\vert \mathcal{A}_i \vert}{\tau+1} + \sum_l \Big( \frac{\nabla^i_{x_{il}} - c}{\vert \mathcal{A}_i \vert} \Big)^{1/\tau} \Big( \frac{\tau\nabla^i_{x_{il}} + c_{\min}}{\tau+1} \Big)
    \\ &= \frac{\vert \mathcal{A}_i \vert}{\tau+1} + \sum_l \Big( \frac{\nabla^i_{x_{il}} - c}{\vert \mathcal{A}_i \vert} \Big)^{1/\tau} \Big( \frac{\tau\nabla^i_{x_{il}} + \max(\nabla^i_{x_i}) - \vert \mathcal{A}_i \vert}{\tau+1} \Big)
    \\ &= \frac{\vert \mathcal{A}_i \vert}{\tau+1} + \sum_l \br_{il} \Big( \frac{\tau\nabla^i_{x_{il}} + \max(\nabla^i_{x_i}) - \vert \mathcal{A}_i \vert}{\tau+1} \Big)
    \\ &= \frac{\vert \mathcal{A}_i \vert}{\tau+1} - \frac{\vert \mathcal{A}_i \vert}{\tau+1} \sum_l \br_{il} + \sum_l \br_{il} \Big( \frac{\tau\nabla^i_{x_{il}} + \max(\nabla^i_{x_i})}{\tau+1} \Big)
    \\ &= \frac{\max(\nabla^i_{x_i})}{\tau+1} \sum_l \br_{il} + \sum_l \br_{il} \Big( \frac{\tau\nabla^i_{x_{il}}}{\tau+1} \Big)
    \\ &= \frac{\max(\nabla^i_{x_i})}{\tau+1} + \frac{\tau}{\tau+1} \br_i^{\tau\top} \nabla^i_{x_i}
    \\ &= \frac{\max(\nabla^i_{x_i})}{\tau+1} + \frac{\tau}{\tau+1} u_i(\br^{\tau}_i, x_{-i})
    \\ &= \frac{1}{1+\tau} [ \max(\nabla^i_{x_i}) + \tau u_i(\br^{\tau}_i, x_{-i}) ]
\end{align}

which implies

\begin{align}
    \max(\nabla^i_{x_i}) - u_i^{\tau}(\br^{\tau}_i, x_{-i}) &\le \tau [ u_i^{\tau}(\br^{\tau}_i, x_{-i}) - u_i(\br^{\tau}_i, x_{-i}) ]
    \\ &= \tau H^{\tau}(\br^{\tau}_i, x_{-i}). \label{eqn:almost_exp}
\end{align}

Recall that $u_i^{\tau}$ is concave in its first argument, and it was proven in previous work~\citep[Lemma 2]{gemp2023approximating} that the following bound holds with respect to the projected gradient:
\begin{align}
    u_i^{\tau}(\br^{\tau}_i, x_{-i}) &\le u_i^{\tau}(\boldsymbol{x}) + \sqrt{2} ||\Pi_{T\Delta}(\nabla^{k\tau}_{x_i})||_2.
\end{align}

We can use this to replace the left $u_i^{\tau}$ term in~\eqref{eqn:almost_exp} and derive a bound on the exploitability:
\begin{align}
    \max(\nabla^i_{x_i}) &\le u_i^{\tau}(\br^{\tau}_i, x_{-i}) + \tau H^{\tau}(\br^{\tau}_i, x_{-i})
    \\ &\le u_i^{\tau}(\boldsymbol{x}) + \tau H^{\tau}(\br^{\tau}_i, x_{-i}) + \sqrt{2} ||\Pi_{T\Delta}(\nabla^{k\tau}_{x_i})||_2
    \\ &= u_i(\boldsymbol{x}) + \underbrace{H^{\tau}(x_i, x_{-i}) + \tau H^{\tau}(\br^{\tau}_i, x_{-i})}_{\text{can measure these if know $\boldsymbol{x}$}} + \sqrt{2} ||\Pi_{T\Delta}(\nabla^{k\tau}_{x_i})||_2
    \\ &\le u_i(\boldsymbol{x}) + (1+\tau) H^{\tau}(\frac{\mathbf{1}}{\vert \mathcal{A}_i \vert}) + \sqrt{2} ||\Pi_{T\Delta}(\nabla^{k\tau}_{x_i})||_2
    \\ &= u_i(\boldsymbol{x}) + \vert \mathcal{A}_i \vert (1 - \sum_l (\frac{1}{\vert \mathcal{A}_i \vert})^{\tau+1}) + \sqrt{2} ||\Pi_{T\Delta}(\nabla^{k\tau}_{x_i})||_2
    \\ &= u_i(\boldsymbol{x}) + \vert \mathcal{A}_i \vert (1 - \vert \mathcal{A}_i \vert^{-\tau}) + \sqrt{2} ||\Pi_{T\Delta}(\nabla^{k\tau}_{x_i})||_2.
\end{align}

This bound is vacuous for payoffs in $(0,1]$ for $\tau=1$ and $\vert \mathcal{A}_i \vert \ge 2$. When is the bound meaningful?
\begin{align}
    \vert \mathcal{A}_i \vert (1 - \vert \mathcal{A}_i \vert^{-\tau}) &\le \epsilon
    \\ 1 - \vert \mathcal{A}_i \vert^{-\tau} &\le \epsilon \vert \mathcal{A}_i \vert^{-1}
    \\ 1 - \epsilon \vert \mathcal{A}_i \vert^{-1} &\le \vert \mathcal{A}_i \vert^{-\tau}
    \\ \log_{\vert \mathcal{A}_i \vert}(1 - \epsilon \vert \mathcal{A}_i \vert^{-1}) &\le -\tau
    \\ \tau &\le -\log_{\vert \mathcal{A}_i \vert}(1 - \epsilon \vert \mathcal{A}_i \vert^{-1})
    \\ &= -\log_{\vert \mathcal{A}_i \vert}(\frac{\vert \mathcal{A}_i \vert - \epsilon}{\vert \mathcal{A}_i \vert})
    \\ &= \log_{\vert \mathcal{A}_i \vert}(\vert \mathcal{A}_i \vert) - \log_{\vert \mathcal{A}_i \vert}(\vert \mathcal{A}_i \vert - \epsilon)
    \\ &= 1 - \log_{\vert \mathcal{A}_i \vert}(\vert \mathcal{A}_i \vert - \epsilon)
\end{align}

Note that $\vert \mathcal{A}_i \vert^{-\tau} = e^{-\ln(\vert \mathcal{A}_i \vert)\tau}$ which is clearly convex in $\tau$. Therefore, $\vert \mathcal{A}_i \vert^{-\tau}$ can be lower bounded about $\tau=0$ by its linear approximation at $\tau=0$:
\begin{align}
    \vert \mathcal{A}_i \vert^{-\tau} &= e^{-\ln(\vert \mathcal{A}_i \vert)\tau}
    \\ &\ge \big( e^{-\ln(\vert \mathcal{A}_i \vert)\tau} \big) \vert_{\tau=0} + \big( -\ln(\vert \mathcal{A}_i \vert) e^{\ln(\vert \mathcal{A}_i \vert)\tau} \big) \vert_{\tau=0} (\tau - 0)
    \\ &= 1 - \ln(\vert \mathcal{A}_i \vert) \tau.
\end{align}

Using this, we can bound the exploitability as a polynomial of $\tau$:
\begin{align}
    \epsilon_i(\boldsymbol{x}) = \max(\nabla^i_{x_i}) - u_i(\boldsymbol{x}) &\le \vert \mathcal{A}_i \vert (1 - \vert \mathcal{A}_i \vert^{-\tau}) + \sqrt{2} ||\Pi_{T\Delta}(\nabla^{k\tau}_{x_i})||_2
    \\ &\le \vert \mathcal{A}_i \vert (1 - 1 + \ln(\vert \mathcal{A}_i \vert) \tau) + \sqrt{2} ||\Pi_{T\Delta}(\nabla^{k\tau}_{x_i})||_2
    \\ &\le \vert \mathcal{A}_i \vert \ln(\vert \mathcal{A}_i \vert) \tau + \sqrt{2} ||\Pi_{T\Delta}(\nabla^{k\tau}_{x_i})||_2.
\end{align}

This means that solving for the Nash equilibrium of the game with $H^{\tau}$ entropy bonuses well approximates the original Nash equilibrium problem at low $\tau$. Recall
\begin{align}
    \nabla_{x_i} u_i^{\tau}(\boldsymbol{x}) &= [I - \frac{1}{\vert \mathcal{A}_i \vert} \mathbf{1} \mathbf{1}^\top] (\nabla^i_{x_i} - \vert \mathcal{A}_i \vert x_i^{\tau}) = 0 \,\, \forall \,\, k
\end{align}
for any equilibrium ($\gamma = \vert \mathcal{A}_i \vert$ ensures all equilibria lie in the interior). Now perform a variable substitution; assume $1/\tau$ is an integer and let $v_i = x_i^{\tau}$. Then $x_i = v_i^{1/\tau}$ and
\begin{align}
    \nabla_{x_i} u_i^{\tau}(\boldsymbol{x}) &= [I - \frac{1}{\vert \mathcal{A}_i \vert} \mathbf{1} \mathbf{1}^\top] (T^i_{1 \ldots n} \ldots v_{k-1}^{1/\tau} v_{k+1}^{1/\tau} \ldots - \vert \mathcal{A}_i \vert v_i) = 0  \,\, \forall \,\, k. \label{app:eqn:zero_proj_grad_poly}
\end{align}

Notice that this is a polynomial of degree $(n-1)(1/\tau)$. In addition, we can include the simplex constraints:
\begin{align}
    \sum_{a_i \in \mathcal{A}_i} v_{i,a_i}^{1/\tau} &= 1
    \\ v_{i,a_i} &\ge 0 \,\, \forall \,\, a_i \in \mathcal{A}_i. \label{app:eqn:nonneg_con}
\end{align}

The entire set of equations represents a multivariate system of polynomial equations. If we want to remove the inequality constraints, we could replace $v_i$ with $z_i$ and let $v_i = z_i^2$.

\section{Tsallis Entropy Bonus as Gumbel Noise}\label{app:gumbel}

It is known that best responses in games with payoffs regularized by Shannon entropy are equivalent to best responding to payoffs perturbed by Gumbel noise $g_i$:
\begin{align}
    \argmax_{i} (g_i + \nabla^i_{x_i} / \tau) &\sim \texttt{Cat}(\texttt{softmax}(\nabla^i_{x_i} / \tau)) = \texttt{Cat}(\frac{\texttt{exp}(\nabla^i_{x_i} / \tau)}{\sum_l \texttt{exp}(\nabla^i_{x_{il}} / \tau)}).
\end{align}
and note
\begin{align}
    \argmax_{i} (\tau g_i + \nabla^i_{x_i}) &= \argmax_{i} (g_i + \nabla^i_{x_i} / \tau)
\end{align}
for $\tau > 0$. Consider the moment generating function for the Gumbel distribution $\tilde{g} \sim $ Gumbel($\tilde{\mu}, \tilde{\beta}$):
\begin{align}
    M_{\tilde{g}}(t) &= \mathbb{E}[e^{t \tilde{g}}] = \Gamma(1 - \tilde{\beta} t) e^{\tilde{\mu} t}.
\end{align}
where $\tilde{\mu} = \mu \tau$ and $\tilde{\beta} = \beta \tau$. By inspection, one can see that this is equivalent to the moment generating function for the distribution $g \sim $ Gumbel($\mu, \beta$):
\begin{align}
    M_g(t) &= \mathbb{E}[e^{t g}] = \Gamma(1 - \beta \tau t) e^{\mu \tau t}.
\end{align}
Therefore, if $g_i \sim$ Gumbel($0, 1$), then a QRE($\tau$) is equivalent to the NE of a game with payoffs perturbed by $g_i \sim$ Gumbel($0, \tau$). In a game with Tsallis entropy bonuses, the best response operator is
\begin{align}
    \br^{\tau}_i &= \frac{(\nabla^i_{x_i})^{1/\tau}}{\sum_l (\nabla^i_{kl})^{1/\tau}} \in \Delta^{\vert \mathcal{A}_i \vert - 1}.
\end{align}
Mapping onto the categorical above, we can see that
\begin{align}
    \argmax_{i} (g_i + \log[(\nabla^i_{x_i})^{1/\tau}]) &\sim \texttt{Cat}(\frac{(\nabla^i_{x_i})^{1/\tau}}{\sum_l (\nabla^i_{kl})^{1/\tau}})
\end{align}
and note
\begin{align}
    \argmax_{i} (g_i + \log[(\nabla^i_{x_i})^{1/\tau}]) &= \argmax_{i} (\tau g_i + \log[\nabla^i_{x_i}])
\end{align}
for $\tau > 0$. The derivation continues as before, but now we see that if $g_i \sim$ Gumbel($0, 1$), then an NE of a game with these precise Tsallis entropy bonuses is equivalent to the NE of a game payoffs are perturbed by $g_i \sim$ Gumbel($0, \tau$) in log-space.

\section{Hyperparameters}\label{app:hyper}

We report the full base configuration of hyperparameters used to solve Chicken below. The parameter $\eta$ represents the learning rate of EigenGame. Skip simply indicates how often to save results.

\begin{verbatim}
def get_config():
  """Get config for EigenGame solver on Chicken game."""
  config = config_dict.ConfigDict()

  config.tau_inv = 3
  config.gamma = 0.5
  config.report_training = True

  config.game_ctor = classic.Chicken
  config.game_config = config_dict.ConfigDict()

  batch_size = 1_000
  hyps = config_dict.ConfigDict()

  hyps.null_space = config_dict.ConfigDict()
  hyps.null_space.k = None
  hyps.null_space.iters = 100_000
  hyps.null_space.eta = 1e-2
  hyps.null_space.batch_size = batch_size
  hyps.null_space.norm_tol = 1e-10
  hyps.null_space.eigenvalue_tolerance = 1e-6
  hyps.null_space.skip = 1_000

  hyps.pinv_shift_1_z = config_dict.ConfigDict()
  hyps.pinv_shift_1_z.k = None
  hyps.pinv_shift_1_z.iters = 10_000
  hyps.pinv_shift_1_z.eta = 1e0
  hyps.pinv_shift_1_z.batch_size = batch_size
  hyps.pinv_shift_1_z.norm_tol = 0.0
  hyps.pinv_shift_1_z.skip = 100

  hyps.pinv_mat_lam = config_dict.ConfigDict()
  hyps.pinv_mat_lam.k = None
  hyps.pinv_mat_lam.iters = 10_000
  hyps.pinv_mat_lam.eta = 1e0
  hyps.pinv_mat_lam.batch_size = batch_size
  hyps.pinv_mat_lam.norm_tol = 0.0
  hyps.pinv_mat_lam.full_eval = False
  hyps.pinv_mat_lam.num_par = 1
  hyps.pinv_mat_lam.skip = 100

  hyps.maxv = config_dict.ConfigDict()
  hyps.maxv.k = 1
  hyps.maxv.iters = 1_000
  hyps.maxv.eta = 1e-1
  hyps.maxv.batch_size = batch_size
  hyps.maxv.norm_tol = 0.0
  hyps.maxv.full_eval = False
  hyps.maxv.num_par = 1
  hyps.maxv.skip = 10

  config.solver_ctor = eigengame.EigenGameSolver
  solver_config = config_dict.ConfigDict()
  solver_config.hyps = hyps
  solver_config.num_lams = 100
  solver_config.sum_to_1_tol = 5e-2
  solver_config.seed = 12345
  config.solver_config = solver_config

  return config
\end{verbatim}

We modify this base configuration depending on the batch size used. The parameter $s$ represents the list of randomly generated seeds used to generate ten thousand random trials.

\begin{verbatim}
num_seeds = 100
s = [int(i) for i in random.randint(999, 9999, size=num_seeds)]
\end{verbatim}

\begin{verbatim}
bs_100 = hyper.product([
    hyper.sweep("config.solver_config.hyps.null_space.skip", [50_000]),
    hyper.sweep("config.solver_config.hyps.pinv_shift_1_z.skip", [50_000]),
    hyper.sweep("config.solver_config.hyps.null_space.batch_size", [100]),
    hyper.sweep("config.solver_config.hyps.pinv_shift_1_z.batch_size", [100]),
    hyper.sweep("config.solver_config.hyps.null_space.iters", [5_000_000]),
    hyper.sweep("config.solver_config.hyps.null_space.eta", [1e-3]),
    hyper.sweep("config.solver_config.hyps.pinv_shift_1_z.iters", [5_000_000]),
    hyper.sweep("config.solver_config.hyps.pinv_shift_1_z.eta", [1e-3]),
    hyper.sweep("config.solver_config.seed", s),
])

bs_200 = hyper.product([
    hyper.sweep("config.solver_config.hyps.null_space.skip", [20_000]),
    hyper.sweep("config.solver_config.hyps.pinv_shift_1_z.skip", [20_000]),
    hyper.sweep("config.solver_config.hyps.null_space.batch_size", [200]),
    hyper.sweep("config.solver_config.hyps.pinv_shift_1_z.batch_size", [200]),
    hyper.sweep("config.solver_config.hyps.null_space.iters", [2_000_000]),
    hyper.sweep("config.solver_config.hyps.null_space.eta", [1e-3]),
    hyper.sweep("config.solver_config.hyps.pinv_shift_1_z.iters", [2_000_000]),
    hyper.sweep("config.solver_config.hyps.pinv_shift_1_z.eta", [1e-3]),
    hyper.sweep("config.solver_config.seed", s),
])

bs_400 = hyper.product([
    hyper.sweep("config.solver_config.hyps.null_space.skip", [10_000]),
    hyper.sweep("config.solver_config.hyps.pinv_shift_1_z.skip", [10_000]),
    hyper.sweep("config.solver_config.hyps.null_space.batch_size", [400]),
    hyper.sweep("config.solver_config.hyps.pinv_shift_1_z.batch_size", [400]),
    hyper.sweep("config.solver_config.hyps.null_space.iters", [100_000]),
    hyper.sweep("config.solver_config.hyps.null_space.eta", [1e-2]),
    hyper.sweep("config.solver_config.hyps.pinv_shift_1_z.iters", [1_000_000]),
    hyper.sweep("config.solver_config.hyps.pinv_shift_1_z.eta", [1e-3]),
    hyper.sweep("config.solver_config.seed", s),
])

bs_500 = hyper.product([
    hyper.sweep("config.solver_config.hyps.null_space.skip", [1_000]),
    hyper.sweep("config.solver_config.hyps.pinv_shift_1_z.skip", [100]),
    hyper.sweep("config.solver_config.hyps.null_space.batch_size", [500]),
    hyper.sweep("config.solver_config.hyps.pinv_shift_1_z.batch_size", [500]),
    hyper.sweep("config.solver_config.hyps.null_space.iters", [100_000]),
    hyper.sweep("config.solver_config.hyps.null_space.eta", [1e0]),
    hyper.sweep("config.solver_config.hyps.pinv_shift_1_z.iters", [1_000]),
    hyper.sweep("config.solver_config.hyps.pinv_shift_1_z.eta", [1e0]),
    hyper.sweep("config.solver_config.seed", s),
])

bs_1000 = hyper.product([
    hyper.sweep("config.solver_config.hyps.null_space.skip", [1_000]),
    hyper.sweep("config.solver_config.hyps.pinv_shift_1_z.skip", [100]),
    hyper.sweep("config.solver_config.hyps.null_space.batch_size", [1_000]),
    hyper.sweep("config.solver_config.hyps.pinv_shift_1_z.batch_size", [1_000]),
    hyper.sweep("config.solver_config.hyps.null_space.iters", [100_000]),
    hyper.sweep("config.solver_config.hyps.null_space.eta", [1e0]),
    hyper.sweep("config.solver_config.hyps.pinv_shift_1_z.iters", [1_000]),
    hyper.sweep("config.solver_config.hyps.pinv_shift_1_z.eta", [1e0]),
    hyper.sweep("config.solver_config.seed", s),
])
\end{verbatim}

\section{Additional Experiments}\label{app:add_exps}

We include additional results on a Bach Stravinsky and Stag Hunt game. These experiments used the same hyperparameters listed in Appendix~\ref{app:hyper}. We found that $\tau^{-1} = 3$ only returned $1$ NE in each game. Increasing to $\tau^{-1}=5$ returned all $3$ NEs using scipy, but we did not run those same experiments for the EigenGame approach we explored.

The Bach Stravinksy game payoffs are
\begin{align}
U^{(1)} &= \begin{bmatrix}
1.0 & 0.01 \\
0.01 & 0.67
\end{bmatrix}
&U^{(2)} = \begin{bmatrix}
0.67 & 0.01 \\
0.01 & 0.67
\end{bmatrix}.
\end{align}

The Stag Hunt payoffs are
Next, we evaluate~\Algref{alg:ne_via_svd} on a Chicken game:
\begin{align}
U^{(1)} &= \begin{bmatrix}
1.0 & 0.01 \\
0.67 & 0.67
\end{bmatrix}
&U^{(2)} = \begin{bmatrix}
1.0 & 0.67 \\
0.01 & 0.67
\end{bmatrix}.
\end{align}

Tables~\ref{tab:bachstravinsky_eigengame} and~\ref{tab:staghunt_eigengame} report the additional results with the ground truth NE for each game (transformed by Tsallis entropy) reported in Tables~\ref{tab:bachstravinsky_gt} and~\ref{tab:staghunt_gt} respectively.

\begin{table}[ht!]
    \centering
    \begin{tabular}{l||c|c|c|c}
        Batch Size & \% Macaulay & \% $S_1 Z$ & Success & Jensen-Shannon \\ \hline\hline
        1000 & 100\% & 100\% & 1.00 & 0.000 \\
        500 & 59\% & 100\% & 1.00 & 0.000 \\
        400 & 48\% & 81\% & 0.90 & 0.002 \\
        200 & 24\% & 40\% & 0.94 & 0.002 \\
        100 & 12\% & 20\% & 0.96 & 0.003
    \end{tabular}
    \caption{(Stochastic Eigendecomposition of Bach Stravinsky with Varying Batch Size) We report the average performance of our approach in approximately recovering the single NE of the Tsallis-transformed Bach Stravinsky game. The first column reports the batch size used in the stochastic SVD and power iteration steps. The second and third translate the batch size into a percentage of the number of rows in the Macaulay ($M$) and $S_1 Z$ matrices respectively ($R_{\lambda}$ contains only $81$ rows which is below our lowest batch size). The Success column reports the success rate of the algorithm returning the one NE. The last column reports the average Jensen-Shannon distance of the returned strategy profile to the ground truth NE of the Tsallis-transformed Bach Stravinsky game. Averages are computed over $100$ trials. The \# of iterations was increased for batch sizes below $400$.}
    \label{tab:bachstravinsky_eigengame}
\end{table}

\begin{table}[ht!]
    \vspace{-0.5cm}
    \centering
    \begin{tabular}{l||c|c|c|c}
        \multirow{2}{*}{$\mathcal{X}^*$} & \multicolumn{2}{c|}{Player 1} & \multicolumn{2}{c}{Player 2} \\
         & $x_{11}$ & $x_{12}$ & $x_{21}$ & $x_{22}$ \\ \hline \hline
        $x^{(1)*}$ & $0.544$ & $0.456$ & $0.456$ & $0.544$ \\
    \end{tabular}
    \caption{(Ground Truth NE of Bach Stravinsky) We use~\Algref{alg:mvp_via_eig} to uncover the true approximate NE for Bach Stravinsky.}
    \label{tab:bachstravinsky_gt}
\end{table}

\begin{table}[ht!]
    \centering
    \begin{tabular}{l||c|c|c|c}
        Batch Size & \% Macaulay & \% $S_1 Z$ & Success & Jensen-Shannon \\ \hline\hline
        1000 & 100\% & 100\% & 0.96 & 0.003 \\
        500 & 59\% & 100\% & 0.96 & 0.003 \\
        400 & 48\% & 81\% & 0.98 & 0.003 \\
        200 & 24\% & 40\% & 0.98 & 0.003 \\
        100 & 12\% & 20\% & 1.00 & 0.003
    \end{tabular}
    \caption{(Stochastic Eigendecomposition of Stag Hunt with Varying Batch Size) We report the average performance of our approach in approximately recovering the single NE of the Tsallis-transformed Stag Hunt game. The first column reports the batch size used in the stochastic SVD and power iteration steps. The second and third translate the batch size into a percentage of the number of rows in the Macaulay ($M$) and $S_1 Z$ matrices respectively ($R_{\lambda}$ contains only $81$ rows which is below our lowest batch size). The Success column reports the success rate of the algorithm returning the one NE. The last column reports the average Jensen-Shannon distance of the returned strategy profile to the ground truth NE of the Tsallis-transformed Stag Hunt game. Averages are computed over $100$ trials. The \# of iterations was increased for batch sizes below $400$.}
    \label{tab:staghunt_eigengame}
\end{table}

\begin{table}[ht!]
    \vspace{-0.5cm}
    \centering
    \begin{tabular}{l||c|c|c|c}
        \multirow{2}{*}{$\mathcal{X}^*$} & \multicolumn{2}{c|}{Player 1} & \multicolumn{2}{c}{Player 2} \\
         & $x_{11}$ & $x_{12}$ & $x_{21}$ & $x_{22}$ \\ \hline \hline
        $x^{(1)*}$ & $0.358$ & $0.642$ & $0.358$ & $0.642$ \\
    \end{tabular}
    \caption{(Ground Truth NE of Stag Hunt) We use~\Algref{alg:mvp_via_eig} to uncover the true approximate NE for Stag Hunt.}
    \label{tab:staghunt_gt}
\end{table}

\end{document}